\documentclass[conference]{IEEEtran}
\IEEEoverridecommandlockouts
\usepackage[letterpaper, left=0.62in, right=0.62in, bottom=1in, top=0.75in]{geometry}
\usepackage{amssymb,amsmath}
\usepackage{graphicx}
\usepackage{cite}
\usepackage{txfonts}
\usepackage{mathrsfs}
\usepackage{fontenc}
\usepackage[binary-units]{siunitx}
\DeclareSIUnit{\bps}{bps}

\usepackage[caption=false,font=footnotesize]{subfig}
\usepackage{tabularx}
\usepackage{multirow}
\usepackage{diagbox}
\usepackage{algorithm2e}
\usepackage{bbm}
\usepackage{xcolor}

\begin{document}

\title{User Selection for Simple Passive Beamforming in Multi-RIS-Aided Multi-User Communications}

\author{\IEEEauthorblockN{Wei Jiang\IEEEauthorrefmark{1}\IEEEauthorrefmark{2} and Hans D. Schotten\IEEEauthorrefmark{2}\IEEEauthorrefmark{1}}
\IEEEauthorblockA{\IEEEauthorrefmark{1}Intelligent Networking Research Group, German Research Center for Artificial Intelligence (DFKI),  Germany\\
  }
\IEEEauthorblockA{\IEEEauthorrefmark{2}Department of Electrical and Computer Engineering, Technische Universit\"at (TU) Kaiserslautern, Germany\\
 }
}

\maketitle

\begin{abstract}
This paper focuses on multi-user downlink signal transmission in a wireless system aided by multiple reconfigurable intelligent surfaces (RISs). In such a multi-RIS, multi-user, multi-antenna scenario, determining a set of RIS phase shifts to maximize the sum throughput becomes intractable.  Hence, we propose a novel scheme that can substantially simplify the optimization of passive beamforming. By opportunistically selecting a user with the best channel condition as the only active transmitter in the system, it degrades to single-user passive beamforming,  where two methods, i.e., joint optimization based on the semidefinite relaxation approach and alternating optimization, are applicable. The superiority of the proposed scheme is demonstrated through Monte-Carlo simulations.
\end{abstract}

\IEEEpeerreviewmaketitle

\section{Introduction}
Reconfigurable intelligent surface (RIS) a.k.a intelligent reflecting surface has attracted growing attention from the research community and is envisaged as a key enabler for the forthcoming sixth-generation (6G) wireless system \cite{Ref_jiang2021road}.
By smartly adjusting the phase shifts of a massive number of reflection elements, RIS enables reconfigurable wireless propagation for particular objectives such as signal amplification and interference suppression. Since each element is passive, low-cost, small, and lightweight, the RIS technology empowers an energy-efficient and cost-efficient way for sustainable capacity and performance growth  \cite{Ref_renzo2020smart}. To exploit its potential, a growing body of literature have studied different aspects for RIS such as joint active and passive beamforming \cite{Ref_wu2019intelligent}, cascaded channel estimation \cite{Ref_liu2020matrix, Ref_zheng2021efficient}, statistical characterization \cite{Ref_do2021multiRIS}, and hardware imperfection \cite{Ref_jiang2023performance}.

Most of the research efforts concentrated on point-to-point RIS communications, consisting of a base station (BS), a single user, and a single RIS, for the sake of analysis. In practice, any wireless system needs to accommodate multiple users simultaneously, imposing some fundamental particularities in the coordination of multi-user signal transmission, as well as the optimization of passive beamforming.  In \cite{Ref_zheng2020intelligent_COML}, the authors compared the performance of typical multiple-access methods in a single-RIS-aided system. Multi-user RIS transmission under discrete phase shifts is studied in \cite{Ref_jiang2022multiuser}. Gan \textit{et al.} \cite{Ref_gan2021user} discussed the impact of multi-user scheduling in RIS-assisted communications.  On the other hand, it is highly possible that a wireless system deploys multiple distributed surfaces, rather than a single RIS. For example, Zheng \textit{et al.} \cite{Ref_zheng2021doubleIRS} aimed to unveil the potential of multi-RIS-aided networks, and the effect of multiple RISs in improving spectral efficiency is studied in \cite{Ref_niu2022double}. The authors of \cite{Ref_jiang2022intelligent} investigated the behaviors of different multi-user schemes in multi-RIS-aided moving vehicular networks.

While the aforementioned works have deepened our understanding, to the best knowledge of the authors, an efficient transmission scheme in a multi-RIS-aided wireless system  is still an open issue. Such a multi-RIS, multi-user, and multi-antenna scenario make the RIS optimization complicated, where determining a set of RIS phase shifts to maximize the sum throughput is intractable.  Responding to this, we aim to propose a novel scheme to simplify passive beamforming. By selecting an opportunistic user with the best channel condition as the anchor,  RIS optimization of a multi-RIS-aided system becomes feasible. Two passive beamforming methods, i.e., joint optimization and alternating optimization, are applicable thanks to this system simplification. Due to the inherent multi-user diversity gain, it outperforms benchmark schemes in terms of achievable sum throughput. Performance evaluation and comparison under \textit{independent but not identically distributed (i.n.i.d.)} Nakagami-\textit{m} fading are conducted to verify the superiority of the proposed scheme.

The remainder of this paper is organized as follows: the next section introduces the system model. Section III presents user selection and two passive beamforming methods. Section IV illustrates simulation results, and conclusions are drawn in Section V.

\section{System Model}

\begin{figure}[!t]
    \centering
    \includegraphics[width=0.425\textwidth]{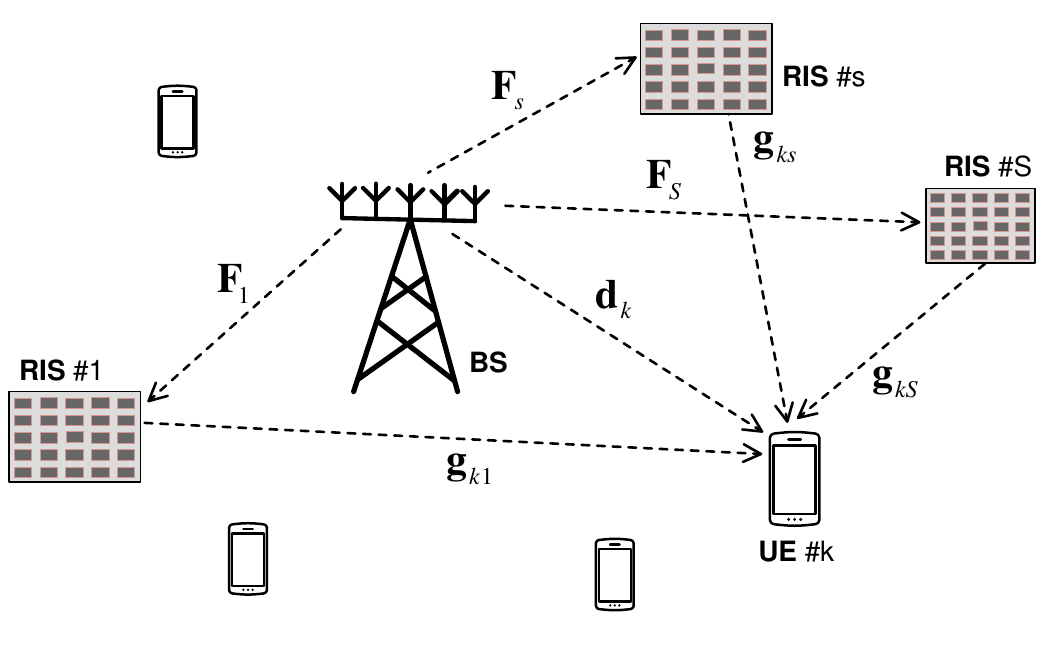}
    \caption{Schematic diagram of a multi-RIS-aided multi-user system.  }
    \label{fig:SystemModel}
\end{figure}

This paper considers the downlink of a wireless system, where $K\geqslant 1$ single-antenna user equipment (UE) access to a $N_b$-antenna BS with the help of $S\geqslant 1$ RIS surfaces, as illustrated in \figurename \ref{fig:SystemModel}. Each surface is equipped with a smart controller that adaptively adjusts the phase shift of each reflecting element based on the knowledge of instantaneous channel state information (CSI).  As most of the prior works \cite{Ref_zheng2020intelligent_COML, Ref_jiang2022multiuser, Ref_gan2021user, Ref_zheng2021doubleIRS, Ref_niu2022double},  the BS is assumed to perfectly know CSI for ease of analysis, and the methods of CSI acquisition refers to the literature such as \cite{Ref_liu2020matrix, Ref_zheng2021efficient}. The $s^{th}$ surface, $\forall s\in \{1,2,\ldots,S\}$, contains $N_s$ elements, and write $M=\sum_{s=1}^SN_s$ denote the total number of elements over all surfaces. Mathematically, a typical element $n_s\in \{1,2,\ldots,N_s\}$ over the $s^{th}$ surface is represented by a reflection coefficient $\epsilon_{sn_s}=e^{j\theta_{sn_s}}$, where $\theta_{sn_s}$ denotes an induced phase shift. Although practical RIS hardware only supports discrete phase shifts, a few phase-control bits (e.g., $2$ bits as argued in \cite{Ref_jiang2022multiuser}) are sufficient for achieving comparable performance as continuous phase shifts. Therefore, the RIS optimization hereinafter utilizes $\theta_{sn_s}\in [0,2\pi)$, $\forall s,n_s$ for notational simplicity.

RIS-aided communications suffer from a heterogeneous fading environment, where mobile users are generally surrounded by sufficient scatters but the BS-RIS link has a line-of-sight (LOS) path since their deployment positions are deliberately selected. Compared to  \textit{independent and identically distributed (i.i.d.)} Rayleigh or Rician channels, generic Nakagami-$m$ fading is more suitable for RIS-aided systems.
Write $ X \sim \mathrm{Nakagami}(m, \Omega)$ to represent a random variable (RV) following Nakagami-$m$ distribution. Its probability
density function (PDF) and cumulative distribution function (CDF) are given by
\begin{align}
    f_X(x)&=\frac{2{m}^{m}}{\Gamma(m){\Omega}^{m}} x^{2m -1} e^{-\frac{m}{\Omega} x^2}\\
    F_X(x)&=\frac{\gamma \left(m, \frac{m}{\Omega} x^2 \right)}{ \Gamma(m)},
\end{align}
where $m > 0$ indicates the severity of small-scale fading, $\Omega > 0$ is equivalent to large-scale fading,  the Gamma function $\Gamma(z)=\int_0^{\infty} t^{z-1}e^{-t}dt$, and the lower incomplete Gamma function $\gamma(z, s)=\int_0^{s} t^{z-1}e^{-t}dt$.  Rayleigh fading is a special case of Nakagami-\textit{m} fading when $m=1$.

Since the fading severity and distance-dependent path losses are distinct for different channels, a more general and practical \textit{i.n.i.d.} setup is employed.
Let $\mathbf{d}_{k}=\left[d_{k1},d_{k2},\ldots,d_{kN_b}\right]^T$ denote the channel vector between  the BS and the $k^{th}$ user. We have $d_{kn_b}=|d_{kn_b}|e^{j\phi^d_{kn_b}}$, $\forall n_b\in \{1,2,\ldots,N_b\}$, where $\phi^d_{kn_b}$ and $|d_{kn_b}|$ stand for the phase and magnitude of a channel coefficient.
The channel between the $s^{th}$ RIS and the $k^{th}$ user is modeled as an $N_s\times 1$ vector $\mathbf{g}_{ks}=\Bigl[g_{k,s1},g_{k,s2},\ldots,g_{k,sN_s}\Bigr]^T$, where $g_{k,sn_s}=|g_{k,sn_s}|e^{j\phi^g_{k,sn_s}}$ is the channel coefficient between the $n_s^{th}$ element of surface $s$ and user $k$. Similarly, write $\textbf{F}_s\in \mathbb{C}^{N_s\times N_b}$ to denote the channel matrix from the BS to the $s^{th}$ surface, where its entry $f_{sn_s,n_b}$ denotes the fading coefficient between BS antenna $n_b$ and the $n_s^{th}$ element of surface $s$. The magnitudes $|d_{kn_b}|$, $|g_{k,sn_s}|$, and $|f_{sn_s,n_b}|$, $\forall s,n_s,k$, are Nakagami-$m$ RVs.

As we know, the BS is capable of simultaneously sending at most $N_b$ information-bearing symbols over the antenna array. Assume there are a total of $K \leqslant N_b$ symbols, where $i_k$ is intended for a general user $k$. The BS utilizes a set of orthogonal beamforming vectors $\mathbf{w}_k\in \mathbb{C}^{N_b\times 1}$, where $\mathbf{w}_k^H \mathbf{w}_{k'}=0$ when $k\neq k'$ and  $\|\mathbf{w}_k\|^2=1$, to superimpose these symbols.  Write $\mathbf{s}\in \mathbb{C}^{N_b\times 1}$ to denote the vector of transmitted signals over the antenna array, we have
\begin{equation} \label{EQnRIS_compositeTxSignal}
    \mathbf{s}=\sum_{k=1}^K\mathbf{w}_k i_k
\end{equation}
with $\mathbb{E}\left[\mathbf{s}^H\mathbf{s}\right]\leqslant 1$. Thus, the $k^{th}$ UE observes the received signal
\begin{equation} \label{eqn_systemModel}
    r_k=\sqrt{P_d}\left(\sum_{s=1}^{S} \sum_{n_s=1}^{N_s} g_{k,sn_s} e^{j\theta_{sn_s}} \mathbf{f}_{sn_s}^T  + \mathbf{d}_{k}^T \right) \mathbf{s} + n_k,
\end{equation}
where $\mathbf{f}_{sn_s}^T\in \mathbb{C}^{ 1\times N_b}$ is the signature of the $n_s^{th}$ element of surface $s$ over the BS antenna array, which corresponds to the $n_s^{th}$ row of $\textbf{F}_s$, $P_d$ expresses the power constraint of the BS, $n_k$ is additive white Gaussian noise (AWGN) with zero mean and variance $\sigma_n^2$, namely $n_k\sim \mathcal{CN}(0,\sigma_n^2)$. Define the diagonal phase-shift matrix for surface $s$ as
\begin{equation}
    \boldsymbol{\Theta}_s=\mathrm{diag}\left\{e^{j\theta_{s1}},\ldots,e^{j\theta_{sN_s}}\right\}, \forall s,
\end{equation} \eqref{eqn_systemModel} can be rewritten in matrix form as
\begin{equation} \label{EQN_IRS_RxSignal_Matrix}
    r_k= \sqrt{P_d}\left(\sum_{s=1}^{S} \mathbf{g}_{ks}^T \boldsymbol{\Theta}_s \mathbf{F}_s +\mathbf{d}_k^T\right)\mathbf{s} +n_k.
\end{equation}

Furthermore, the channel vector from all reflecting elements to the $k^{th}$ user is denoted by
\begin{equation}
    \mathbf{g}_k\in \mathbb{C}^{M\times 1}=\left[\mathbf{g}_{k1}^T,\mathbf{g}_{k2}^T,\ldots,\mathbf{g}_{kS}^T \right]^T,
\end{equation}
and the channel matrix from the BS to all reflecting elements is given by
\begin{equation}
\mathbf{F}\in \mathbb{C}^{M\times N_b}=\left[\mathbf{F}_{1}^T,\mathbf{F}_{2}^T,\ldots,\mathbf{F}_{S}^T \right]^T.
\end{equation}
Defining an overall phase-shift matrix
\begin{equation}
    \boldsymbol{\Theta}=\mathrm{diag}\biggl\{\boldsymbol{\Theta}_1,\boldsymbol{\Theta}_2,\ldots,\boldsymbol{\Theta}_S\biggr\},
\end{equation}
\eqref{EQN_IRS_RxSignal_Matrix} is further simplified to
\begin{equation} \label{EQN_RIS_matrixSignalModel}
    r_k= \sqrt{P_d}\left(\mathbf{g}_{k}^T \boldsymbol{\Theta} \mathbf{F} +\mathbf{d}_k^T\right)\mathbf{s} +n_k.
\end{equation}
Assuming the $k^{th}$ user knows the CSI, its desired symbol can be detected properly. For example, applying matched filtering or maximal-ratio combining yields
\begin{align} \label{EQN_RIS_RxSignalDetection} \nonumber
  \hat{i}_k &= \frac{\left[\left(\mathbf{g}_{k}^T \boldsymbol{\Theta} \mathbf{F} +\mathbf{d}_k^T\right)\mathbf{w}_k\right]^H}{\left|\left(\mathbf{g}_{k}^T \boldsymbol{\Theta} \mathbf{F} +\mathbf{d}_k^T\right)\mathbf{w}_k \right|} r_k\\ \nonumber &= \frac{\left[\left(\mathbf{g}_{k}^T \boldsymbol{\Theta} \mathbf{F} +\mathbf{d}_k^T\right)\mathbf{w}_k\right]^H}{\left|\left(\mathbf{g}_{k}^T \boldsymbol{\Theta} \mathbf{F} +\mathbf{d}_k^T\right)\mathbf{w}_k \right|} \left(\sqrt{P_d}\left(\mathbf{g}_{k}^T \boldsymbol{\Theta} \mathbf{F} +\mathbf{d}_k^T\right) \sum_{k}\mathbf{w}_ki_k +n_k\right)\\
  &= \sqrt{P_d}\left|\left(\mathbf{g}_{k}^T \boldsymbol{\Theta} \mathbf{F} +\mathbf{d}_k^T\right)\mathbf{w}_k \right| i_k +n_k',
\end{align}
since $\mathbf{w}_k^H \mathbf{w}_{k'}=0$ when $k\neq k'$.

\section{User Selection-Based RIS Transmission and Reflection}
In a point-to-point system, the channel capacity provides a measure of the performance limit: reliable communications with an arbitrarily small error probability can be achieved at any rate $R<C$, whereas reliable communications are impossible when $R>C$. For a multi-user system consisting of a BS and $K$ users, the concept is extended to a similar performance metric called \textit{a capacity region} \cite{Ref_tse2005fundamentals}. It is characterized by a $K$-dimensional space $\mathfrak{C} \in \mathbb{R}_+^K$, where $\mathbb{R}_+$ denotes the set of non-negative real-valued numbers, and $\mathfrak{C}$ is the set of all K-tuples $(R_1,R_2,\ldots,R_{K})$ such that a generic user $k$ can reliably communicate at rate $R_k$ simultaneously with others. Due to the shared transmission resource, there is a trade-off: if one desires a higher rate, some of other users have to lower their rates. From this capacity region, a performance metric can be derived, i.e., the sum capacity
\begin{equation}
  C= \max_{(R_1,R_2,\ldots,R_{K})\in \mathfrak{C}} \left( \sum_{k=1}^K R_k   \right),
\end{equation}
indicating the maximum total throughput that can be achieved.

The aim of this paper is to maximize the sum capacity of a multi-RIS-aided system, namely
\begin{equation} \label{EQN_SumRate}
    C =\sum_{k=1}^K \log\left(1+\frac{ \Bigl| \left(\mathbf{g}_{k}^T \boldsymbol{\Theta} \mathbf{F} +\mathbf{d}_k^T\right)\mathbf{w}_k  \Bigr|^2 P_d}{\sigma_n^2} \right),
\end{equation}
recalling \eqref{EQN_RIS_RxSignalDetection}.
It formulates the following optimization:
\begin{equation}
\begin{aligned} \label{eqnIRS:optimizationMRTvector}
\max_{\boldsymbol{\Theta},\:\mathbf{w}_k}\quad &  \sum_{k=1}^K \log\left(1+\frac{ \Bigl| \left(\mathbf{g}_{k}^T \boldsymbol{\Theta} \mathbf{F} +\mathbf{d}_k^T\right)\mathbf{w}_k  \Bigr|^2 P_d}{\sigma_n^2} \right) \\
\textrm{s.t.} \quad & \|\mathbf{w}_k\|^2= 1,\: \forall k\\
  \quad & \theta_{sn_s}\in [0,2\pi), \: \forall s,n_s,
\end{aligned}
\end{equation}
which is non-convex because the objective function is not jointly concave with respect to $\boldsymbol{\Theta}$ and $\mathbf{w}_k$. To the best knowledge of the authors, \eqref{eqnIRS:optimizationMRTvector} is intractable, and therefore the optimization of passive beamforming is infeasible.
Hence, it is necessary to explore an innovative way to simplify the system design in such a multi-RIS-aided multi-user system.

In wireless systems, opportunistic communications \cite{Ref_jiang2016robust} achieve a well performance-complexity trade-off by exploiting multi-user diversity gain. Inspired by this, we propose a novel scheme that can simplify passive beamforming. By selecting an opportunistic user with the best channel condition as the anchor, a multi-user system falls back to a single-user system, where RIS optimization is simplified and becomes feasible.
The \textit{single-user bound} for a typical user $k$, which means the capacity of a point-to-point link with the other users absent from the system,  is given by
\begin{equation} \label{EQN_IRS_SingleUserBound}
  R_k = \log  \left(  1+\Bigl| \left(\mathbf{g}_{k}^T \boldsymbol{\Theta} \mathbf{F} +\mathbf{d}_k^T\right)\mathbf{w}_k  \Bigr|^2\frac{P_d}{\sigma_n^2}  \right),\: \forall\: k.
\end{equation}

The philosophy of user selection is to determine the best user achieving the largest rate, i.e.,
\begin{equation}
    k^\star=\arg \max_{k\in\{1,\ldots,K\}} \left\{R_k\right\}.
\end{equation}
The BS only transmits the information-bearing symbol towards $k^\star$, such that the transmitted signal vector in  \eqref{EQnRIS_compositeTxSignal} becomes $\mathbf{s}=\mathbf{w}_{k^\star}i_{k^\star}$. Without inter-user interference, the linear beamforming of maximal-ratio transmission (MRT) that maximizes the strength of the desired signal is optimal. That is
\begin{equation}  \label{RISEQN_MRTvector}
    \mathbf{w}_{k^\star} = \frac{\left(\mathbf{g}_{k^\star}^T \boldsymbol{\Theta} \mathbf{F} +\mathbf{d}_{k^\star}^T\right)^H}{\left\|\mathbf{g}_{k^\star}^T \boldsymbol{\Theta} \mathbf{F} +\mathbf{d}_{k^\star}^T\right\|},
\end{equation} where $\| \cdot\|$ expresses the Frobenius norm. Using $\mathbf{w}_{k^\star}$ and  $\mathbf{w}_k=\mathbf{0}$, when $k\neq k^\star$ into \eqref{EQN_SumRate} yields the sum throughput
\begin{equation} \label{EQN_IRS_MTRcapacity}
  C = \log  \left(  1+\Bigl\| \mathbf{g}_{k^\star}^T \boldsymbol{\Theta} \mathbf{F} +\mathbf{d}_{k^\star}^T  \Bigr\|^2 \frac{P_d}{\sigma_n^2}  \right).
\end{equation} Correspondingly, \eqref{eqnIRS:optimizationMRTvector} is simplified to
\begin{equation}
\begin{aligned} \label{EQNRIS_optimizationUS}
\max_{\boldsymbol{\Theta}}\quad &    \Bigl\| \mathbf{g}_{k^\star}^T \boldsymbol{\Theta} \mathbf{F} +\mathbf{d}_{k^\star}^T  \Bigr\|^2  \\
\textrm{s.t.} \quad &\theta_{sn_s}\in [0,2\pi), \: \forall s,n_s.
\end{aligned}
\end{equation}

For a typical user $k$, we have
\begin{equation} \label{EQNRIS_powerGainE2E}
    \Bigl\| \mathbf{g}_{k}^T \boldsymbol{\Theta} \mathbf{F} +\mathbf{d}_k^T  \Bigr\|^2= \sum_{n_b=1}^{N_b} \left|  \sum_{s=1}^{S} \sum_{n_s=1}^{N_s}  g_{k,sn_s} e^{j\theta_{sn_s}} f_{sn_s,n_b} + d_{kn_b}  \right|^2.
\end{equation}
Ideally, if not consider the implementation limit, we can derive the theoretical maximum of \eqref{EQNRIS_powerGainE2E} as
\begin{equation} \label{EQN_RIS_GammaK}
   \gamma_k^{Max} =  \sum_{n_b=1}^{N_b} \left| \sum_{s=1}^{S} \sum_{n_s=1}^{N_s}  | g_{k,sn_s}||f_{sn_s,n_b} |  + |d_{kn_b}|  \right|^2.
\end{equation}

The optimal phase shifts to achieve \eqref{EQN_RIS_GammaK} equal $\theta_{sn_s}= \phi^d_{kn_b}-(\phi^g_{k,sn_s}+\phi^f_{sn_s,n_b})$. Since both $\phi^d_{k,n_b}$ and $\phi^f_{sn_s,n_b}$ are different for different BS antennae,  each reflecting element needs to provide antenna-specific phase shifts. However, each element can only be tuned to a single phase-shifting value at any time. Hence, the optimal value given in \eqref{EQN_RIS_GammaK} is not achievable. We need to find a sub-optimal solution to obtain a sum throughput as large as possible, which will be presented in the following two sub-sections.
The proposed scheme is depicted in Algorithm 1.
\SetKwComment{Comment}{/* }{ */}
\RestyleAlgo{ruled}
\begin{algorithm}
\caption{User Selection in RIS Communications} \label{alg:IRS002}
\ForEach{Coherence Interval}{
Estimate $\mathbf{F}_s$, $\mathbf{d}_k$, and $\mathbf{g}_{sk}$, $\forall s,k$\;
$\gamma_k^{Max} \gets  \sum_{n_b=1}^{N_b} \left| \sum_{s=1}^{S} \sum_{n_s=1}^{N_s}  | g_{k,sn_s}||f_{sn_s,n_b} |  + |d_{kn_b}|  \right|^2$\;
$k^\star \gets \arg \max_{k\in\{1,\ldots,K\}} \bigl( \gamma_k^{Max} \bigr)$\;
Calculate $\mathbf{\Theta}_{jo}$ through SDR or $\mathbf{\Theta}_{ao}$ via AO\;
Tune RISs to $\mathbf{\Theta}_{jo}$ or $\mathbf{\Theta}_{ao}$\;
The BS transmits $\mathbf{s}=\mathbf{w}_{k^\star}i_{k^\star}$\;
}
\end{algorithm}

\subsection{Joint Optimization (JO)}
In contrast to the \textit{optimal} phase shifts, a set of \textit{optimized} phase shifts can be obtained by solving \eqref{EQNRIS_optimizationUS}, which is a non-convex quadratically constrained quadratic program (QCQP). As we know, the semidefinite relaxation (SDR) approach is effective to solve a QCQP \cite{Ref_wu2019intelligent}.
Define $\mathbf{q}=\left[q_{1},q_{2},\ldots,q_{M}\right]^H$, where $q_{m}=e^{j\theta_{sn_s}}$ with $m=\sum_{s'=1}^{s-1}N_{s'}+n_s$. Let  $\boldsymbol{\chi}=\mathrm{diag}(\mathbf{g}_{k^\star}^T) \mathbf{F} \in \mathbb{C}^{M\times N_b}$, we have $\mathbf{g}_{k^\star}^T \boldsymbol{\Theta} \mathbf{F} =\mathbf{q}^H\boldsymbol{\chi}\in \mathbb{C}^{1\times N_b} $.
Thus, $\left\|\mathbf{g}_{k^\star}^T \boldsymbol{\Theta} \mathbf{F}  +\mathbf{d}_{k^\star}^T\right\|^2=\left\|\mathbf{q}^H\boldsymbol{\chi} +\mathbf{d}_{k^\star}^T\right\|^2$.
 Introducing an auxiliary variable $t$, \eqref{EQNRIS_optimizationUS} is homogenized as
\begin{equation} \begin{aligned} \label{eqn_IRS_relaxedOptimization}
    \max_{\mathbf{q}}  \quad & \left\|t\mathbf{q}^H\boldsymbol{\chi} +\mathbf{d}_{k^\star}^T\right\|^2\\
     =\max_{\mathbf{q}} \quad & t^2\mathbf{q}^H\boldsymbol{\chi}\boldsymbol{\chi}^H\mathbf{q}+t\mathbf{q}^H\boldsymbol{\chi}\mathbf{d}_{k^\star}^*+t\mathbf{d}_{k^\star}^T\boldsymbol{\chi}^H\mathbf{q}+\|\mathbf{d}_{k^\star}\|^2.
     \end{aligned}
\end{equation}
Define $\mathbf{C}=\begin{bmatrix}\boldsymbol{\chi}\boldsymbol{\chi}^H&\boldsymbol{\chi}\mathbf{d}_{k^\star}^H\\ \mathbf{d}_{k^\star}\boldsymbol{\chi}^H& \|\mathbf{d}_{k^\star}\|^2\end{bmatrix},\:\:\mathbf{v}= \begin{bmatrix}\mathbf{q}\\ t\end{bmatrix}$, and $\mathbf{V}=\mathbf{v}\mathbf{v}^H$, we have $\mathbf{v}^H\mathbf{C}\mathbf{v}=\mathrm{Tr}(\mathbf{C}\mathbf{V})$, where $\mathrm{Tr}(\cdot)$ denotes the trace of a matrix.
Thus, \eqref{eqn_IRS_relaxedOptimization} is reformulated as
\begin{equation}  \label{RIS_EQN_optimizationTrace}
\begin{aligned} \max_{\mathbf{V}}\quad &  \mathrm{Tr} \left(  \mathbf{C}\mathbf{V} \right)\\
\textrm{s.t.}  \quad & \mathbf{V}_{m,m}=1, \: \forall m=1,\ldots,M \\
  \quad & \mathbf{V}\succ 1
\end{aligned},
\end{equation}
where $\mathbf{V}_{m,m}$ means the $m^{th}$ diagonal element of $\mathbf{V}$, and $\succ$ stands for a positive semi-definite matrix.
The optimization finally becomes a semi-definite program, whose globally optimal solution $\mathbf{V}^\star$ can be got by a numerical algorithm named CVX \cite{cvx}.

A sub-optimal solution for \eqref{RIS_EQN_optimizationTrace} is given by $\bar{\mathbf{v}}=\mathbf{U}\boldsymbol{\Sigma}^{1/2}\mathbf{r}$,
where $\mathbf{r}\in \mathcal{CN}(\mathbf{0},\mathbf{I}_{M+1})$ is a Gaussian RV, a unitary matrix $\mathbf{U}$ and a diagonal matrix $\boldsymbol{\Sigma}$ are obtained from the eigenvalue decomposition $\mathbf{V}^\star=\mathbf{U}\boldsymbol{\Sigma}\mathbf{U}^H$. The jointly optimized phase shifts are determined as $\boldsymbol{\Theta}_{jo} =\mathrm{diag} \left\{e^{j \arg\left( \left[\frac{\bar{\mathbf{v}}}{\bar{v}_{_{M+1}}} \right]_{1:M}\right)}\right\}$, where $[\cdot]_{1:M}$ denotes a sub-vector extracting the first $M$ elements, $\bar{v}_{_{M+1}}$ is the last element of $\bar{\mathbf{v}}$, and $\arg(\cdot)$ represents the angle of a complex vector or scalar. Applying $k=k^\star$ and $\boldsymbol{\Theta}_{jo}$ into \eqref{EQN_IRS_MTRcapacity} yields a sum capacity
\begin{equation}
  C = \log  \left(  1+\Bigl\| \mathbf{g}_{k^\star}^T \boldsymbol{\Theta}_{jo} \mathbf{F} +\mathbf{d}_{k^\star}^T  \Bigr\|^2 \frac{P_d}{\sigma_n^2}  \right).
\end{equation}

\subsection{Alternating Optimization (AO)}
Although SDR can figure out optimized phase shifts, it suffers from prohibitive computational complexity, blocking its use in practical systems.
To lower the complexity, we can apply alternating optimization that alternately optimizes $\boldsymbol{\Theta}$ and $\mathbf{w}_k$ in an iterative manner \cite{Ref_wu2019intelligent}.
Without loss of generality, the MRT for the direct link can be applied as the initial value of the transmit vector, i.e., $\mathbf{w}_{k^\star}^{(0)}=\mathbf{d}_{k^\star}^*/\|\mathbf{d}_{k^\star}\|$.
Thus, \eqref{eqnIRS:optimizationMRTvector} is simplified to
\begin{equation}  \label{eqnIRS:optimAO}
\begin{aligned} \max_{\boldsymbol{\Theta}}\quad &  \biggl|\Bigl(\mathbf{g}_{k^\star}^T \boldsymbol{\Theta} \mathbf{F} +\mathbf{d}_{k^\star}^T\Bigr)\mathbf{w}_{k^\star}^{(0)}\biggr|^2\\
\textrm{s.t.}  \quad & \theta_{sn_s}\in [0,2\pi), \: \forall s,n_s.
\end{aligned}
\end{equation}
The objective function is still non-convex but it enables a closed-form solution by applying the well-known triangle inequality
\begin{equation}
    \biggl| \Bigl(\mathbf{g}_{k^\star}^T \boldsymbol{\Theta} \mathbf{F} +\mathbf{d}_{k^\star}^T\Bigr)\mathbf{w}_{k^\star}^{(0)} \biggr| \leqslant \biggl|\mathbf{g}_{k^\star}^T \boldsymbol{\Theta} \mathbf{F}\mathbf{w}_{k^\star}^{(0)} \biggr| +\biggl|\mathbf{d}_{k^\star}^T\mathbf{w}_{k^\star}^{(0)}\biggr|.
\end{equation}
Equality achieves if and only if
\begin{equation}
    \arg\left (\mathbf{g}_{k^\star}^T \boldsymbol{\Theta} \mathbf{F}\mathbf{w}_{k^\star}^{(0)} \right)= \arg\left(\mathbf{d}_{k^\star}^T\mathbf{w}_{k^\star}^{(0)}\right)\triangleq \varphi_{0}.
\end{equation}

Recall $\mathbf{q}=\left[q_{1},q_{2},\ldots,q_{M}\right]^H$  and define $\boldsymbol{x}=\mathrm{diag}(\mathbf{g}_{k^\star}^T )\boldsymbol{\Theta} \mathbf{F}\mathbf{w}_{k^\star}^{(0)}\in \mathbb{C}^{M\times 1}$, we have $\mathbf{g}_{k^\star}^T \boldsymbol{\Theta} \mathbf{F}\mathbf{w}_{k^\star}^{(0)}=\mathbf{q}^H\boldsymbol{x}\in \mathbb{C} $.
Ignore the constant term $\bigl|\mathbf{d}_{k^\star}^T\mathbf{w}_{k^\star}^{(0)}\bigr|$, \eqref{eqnIRS:optimAO} is transformed to
\begin{equation}  \label{eqnIRS:optimizationQ}
\begin{aligned} \max_{\boldsymbol{\mathbf{q}}}\quad &  \Bigl|\mathbf{q}^H\boldsymbol{x}\Bigl|\\
\textrm{s.t.}  \quad & |q_{m}|=1, \: \forall m,\\
  \quad & \arg(\mathbf{q}^H\boldsymbol{x})=\varphi_{0}.
\end{aligned}
\end{equation}
The solution for \eqref{eqnIRS:optimizationQ} can be derived as
\begin{equation} \label{eqnIRScomplexityQ}
    \mathbf{q}^{(1)}=e^{j\left(\varphi_{0}-\arg(\boldsymbol{x})\right)}=e^{j\left(\varphi_{0}-\arg\left( \mathrm{diag}(\mathbf{g}_{k^\star}^T )\mathbf{F}\mathbf{w}_{k^\star}^{(0)}\right)\right)}.
\end{equation}
Accordingly,
\begin{align} \nonumber \label{IRSeqnOptimalShift}
    \theta_{sn_s}^{(1)}&=\varphi_{0}-\arg\left(g_{k^\star,sn_s} \mathbf{f}_{sn_s}^T\mathbf{w}_{k^\star}^{(0)}\right)\\&=\varphi_{0}-\arg\left(g_{k^\star,sn_s}\right)-\arg\left(\mathbf{f}_{sn_s}^T\mathbf{w}_{k^\star}^{(0)}\right),
\end{align}
where $\mathbf{f}_{sn_s}^T\mathbf{w}_{k^\star}^{(0)}\in \mathbb{C} $ can be regarded as an effective channel perceived by the $n^{th}$ reflecting element, combining the effects of transmit beamforming $ \mathbf{w}_{k^\star}^{(0)}$ and channel response $\mathbf{f}_{sn_s}$.
Once the optimized phase-shift matrix at the first iteration denoted by $    \boldsymbol{\Theta}^{(1)}=\mathrm{diag}\left\{e^{j\theta_{11}^{(1)}},e^{j\theta_{12}^{(1)}},\ldots,e^{j\theta_{SN_S}^{(1)}}\right\}$ is known, the process alternates to optimize $\mathbf{w}_{k^\star}$. The BS can apply MRT to maximize the strength of the desired signal, resulting in
\begin{equation}  \label{EQN_IRS_TXBF}
    \mathbf{w}_{k^\star}^{(1)} = \frac{  \left(\mathbf{g}_{k^\star}^T \boldsymbol{\Theta}^{(1)} \mathbf{F} +\mathbf{d}_{k^\star}^T\right)^H }{\left\|  \mathbf{g}_{k^\star}^T \boldsymbol{\Theta}^{(1)} \mathbf{F} +\mathbf{d}_{k^\star}^T \right\|}.
\end{equation}

Once the first iteration completes, the BS knows $\boldsymbol{\Theta}^{(1)}$ and $\mathbf{w}_{k^\star}^{(1)}$, which serve as the input for the second iteration for further calculating $\boldsymbol{\Theta}^{(2)}$ and $\mathbf{w}_{k^\star}^{(2)}$.
This process iterates until the convergence is achieved with the optimal beamformer $\mathbf{w}_{k^\star}^{\star}$ and  optimal reflection $\boldsymbol{\Theta}_{ao}$. Substituting $\mathbf{w}_{k^\star}^{\star}$ and  $\boldsymbol{\Theta}_{ao}$ into \eqref{EQN_IRS_MTRcapacity}, we gets the sum capacity, which is given by
\begin{equation}
  C = \log  \left(  1+\Bigl\| \mathbf{g}_{k^\star}^T \boldsymbol{\Theta}_{ao} \mathbf{F} +\mathbf{d}_{k^\star}^T  \Bigr\|^2 \frac{P_d}{\sigma_n^2}  \right).
\end{equation}

\section{Numerical results}

Monte-Carlo simulations are conducted to comparatively evaluate the performance of different schemes in a multi-RIS-aided system.
Without loss of generality, we established the following simulation setup: an 8-antenna BS is located at the center of a circular cell with the radius of \SI{300}{\meter}, where $K=4$ users are  randomly distributed. Four surfaces with $N_s=200$ elements each are deployed equally on a concentric circle with a distance of \SI{90}{\meter} to the BS. The BS transmit power is set to $P_d=20\mathrm{W}$ following the specifications of 3GPP. Signal bandwidth is $10\mathrm{MHz}$ and the noise power density equals $-174\mathrm{dBm/Hz}$ with noise figure  $9\mathrm{dB}$.
The large-scale fading for non-line-of-sight condition is computed according to the 3GPP Urban Micro (UMi) model as $\Omega = - 22.7 - 26 \log(f_c) - 36.7 \log(d)$, where $d$ is the distance and carrier frequency $f_c=2\mathrm{GHz}$. Considering the LOS component, the path loss of the BS-RIS link is calculated by $\mathcal{L}_0/d^{-\alpha}$,
where $\mathcal{L}_0=\SI{-30}{\decibel}$ is the path loss at the reference distance of \SI{1}{\meter} and the path-loss exponent $\alpha=2$. For small-scale fading, the severity of all Nakagami-m channels is set to $m=2.5$.

\begin{figure}[!t]
    \centering
    \includegraphics[width=0.38\textwidth]{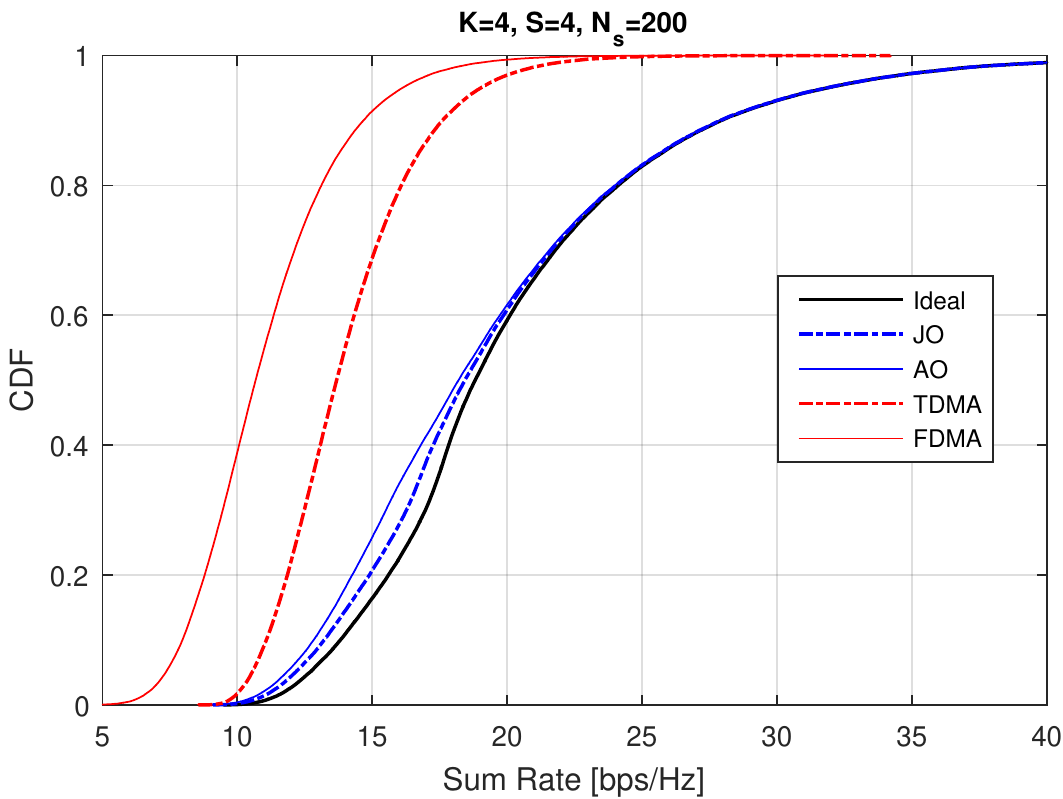}
    \caption{Performance comparison of different transmission schemes in a multi-RIS-aided multi-user system.  }
    \label{Fig_outagePerformance}
\end{figure}

We employ the cumulative distribution function (CDF) in terms of achievable sum throughput as the performance measurement. The transmission schemes for comparison include 1) Time-division multiple access (TDMA); 2) Frequency-division multiple access (FDMA); 3) User selection with joint optimization (JO); 4) User selection with alternating optimization (AO), where the number of optimization iterations is three, which is sufficient for convergence; and 5) User selection with ideal optimization, as depicted by \eqref{EQN_RIS_GammaK}.
As shown in \figurename \ref{Fig_outagePerformance}, TDMA is better than FDMA because the RIS surfaces can provide time-selective reflection particularly optimized for each TDMA user, whereas  the FDMA users suffer from the lack of frequency-selective reflection. However, TDMA and FDMA merely use $1/K$ of the total resource, leading to a rate loss according to the Shannon theorem. The proposed scheme fully exploits the inherent multi-user gain via the user selection, where only the user on the best channel condition utilizes the whole time-frequency resource, resulting in a remarkable capacity gain. Both joint and alternating optimization achieve a near-optimal performance indicated by the ideal optimization.

Furthermore, the complexity is evaluated in terms of average CPU run time per channel realization. The platform uses an Intel i7-4790 CPU at 3.60GHz and 32GB random access memory and runs Matlab parallel computing with 4 workers. As shown in Table I, due to the use of the computation-hungry SDR approach, the complexity of JO is three orders of magnitude higher than that of the benchmark schemes. Luckily, the alternative AO scheme can substantially lower the complexity, which is comparable to TDMA and FDMA. Considering the channel coherence time generally falls into the range of \SIrange{10}{100}{\milli \second}, JO is prohibitive for practical uses but AO is quite efficient.
\begin{table}[!h]
\renewcommand{\arraystretch}{1.3}
\caption{Comparison of computational complexity.}
\label{table_complexity}
\centering
\begin{tabular}{|c|c|c|c|c|}
\hline
Schemes & \textit{JO} & \textit{AO}&\textit{TDMA}& \textit{FDMA} \\
\hline \hline
CPU time [\si{\milli \second}] & $5.47\times 10^4$&  24 &    92 & 22  \\ \hline
\end{tabular}
\end{table}

\section{Conclusions}
In this paper, we proposed a novel transmission scheme that can simplify the optimization of passive beamforming in a multi-RIS-aided multi-user, multi-antenna system. By selecting an opportunistic user with the best channel condition as the anchor,  the RIS optimization of a multi-RIS-aided system becomes feasible through two proposed methods, i.e., joint optimization and alternating optimization. Thanks to the inherent multi-user diversity gain, it outperforms FDMA and TDMA in terms of achievable sum throughput. Although joint optimization is prohibitive for practical uses due to its high complexity, alternating optimization is simple for implementation.
\bibliographystyle{IEEEtran}
\bibliography{IEEEabrv,Ref_VTC}

\begin{thebibliography}{10}
\providecommand{\url}[1]{#1}
\csname url@samestyle\endcsname
\providecommand{\newblock}{\relax}
\providecommand{\bibinfo}[2]{#2}
\providecommand{\BIBentrySTDinterwordspacing}{\spaceskip=0pt\relax}
\providecommand{\BIBentryALTinterwordstretchfactor}{4}
\providecommand{\BIBentryALTinterwordspacing}{\spaceskip=\fontdimen2\font plus
\BIBentryALTinterwordstretchfactor\fontdimen3\font minus
  \fontdimen4\font\relax}
\providecommand{\BIBforeignlanguage}[2]{{%
\expandafter\ifx\csname l@#1\endcsname\relax
\typeout{** WARNING: IEEEtran.bst: No hyphenation pattern has been}%
\typeout{** loaded for the language `#1'. Using the pattern for}%
\typeout{** the default language instead.}%
\else
\language=\csname l@#1\endcsname
\fi
#2}}
\providecommand{\BIBdecl}{\relax}
\BIBdecl

\bibitem{Ref_jiang2021road}
W.~Jiang \emph{et~al.}, ``The road towards {6G}: A comprehensive survey,''
  \emph{IEEE Open J. Commun. Society}, vol.~2, pp. 334--366, Feb. 2021.

\bibitem{Ref_renzo2020smart}
M.~D. Renzo \emph{et~al.}, ``Smart radio environments empowered by
  reconfigurable intelligent surfaces: How it works, state of research, and the
  road ahead,'' \emph{{IEEE} J. Sel. Areas Commun.}, vol.~38, no.~11, pp. 2450
  -- 2525, Nov. 2020.

\bibitem{Ref_wu2019intelligent}
Q.~Wu and R.~Zhang, ``Intelligent reflecting surface enhanced wireless network
  via joint active and passive beamforming,'' \emph{{IEEE} Trans. Wireless
  Commun.}, vol.~18, no.~11, pp. 5394 -- 5409, Nov. 2019.

\bibitem{Ref_liu2020matrix}
H.~Liu, X.~Yuan, and Y.-J.~A. Zhang, ``Matrix-calibration-based cascaded
  channel estimation for reconfigurable intelligent surface assisted multiuser
  {MIMO},'' \emph{{IEEE} J. Sel. Areas Commun.}, vol.~38, no.~11, pp. 2621 --
  2636, Nov. 2020.

\bibitem{Ref_zheng2021efficient}
B.~Zheng, C.~You, and R.~Zhang, ``Efficient channel estimation for double-irs
  aided multi-user {MIMO} system,'' \emph{{IEEE} Trans. Commun.}, vol.~69,
  no.~6, pp. 3818 -- 3832, Jun. 2021.

\bibitem{Ref_do2021multiRIS}
T.~N. Do \emph{et~al.}, ``Multi-{RIS}-aided wireless systems: Statistical
  characterization and performance analysis,'' \emph{{IEEE} Trans. Commun.},
  vol.~69, no.~12, pp. 8641 -- 8658, Dec. 2021.

\bibitem{Ref_jiang2023performance}
W.~Jiang and H.~Schotten, ``Performance impact of channel aging and phase noise
  on intelligent reflecting surface,'' \emph{{IEEE} Commun. Lett.}, 2022, early
  Access.

\bibitem{Ref_zheng2020intelligent_COML}
B.~Zheng, Q.~Wu, and R.~Zhang, ``Intelligent reflecting surface-assisted
  multiple access with user pairing: {NOMA or OMA}?'' \emph{{IEEE} Commun.
  Lett.}, vol.~24, no.~4, pp. 753 -- 757, Apr. 2020.

\bibitem{Ref_jiang2022multiuser}
W.~Jiang and H.~Schotten, ``Multi-user reconfigurable intelligent surface-aided
  communications under discrete phase shifts,'' in \emph{Proc. 36th IEEE Int.
  Workshop on Commun. Qual. and Reliability (CQR 2022)}, Arlington, United
  States, Sep. 2022.

\bibitem{Ref_gan2021user}
X.~Gan \emph{et~al.}, ``User selection in reconfigurable intelligent surface
  assisted communication systems,'' \emph{{IEEE} Commun. Lett.}, vol.~25,
  no.~4, pp. 1353 -- 1357, Apr. 2021.

\bibitem{Ref_zheng2021doubleIRS}
B.~Zheng, C.~You, and R.~Zhang, ``Double-{IRS} assisted multi-user {MIMO}:
  Cooperative passive beamforming design,'' \emph{{IEEE} Trans. Wireless
  Commun.}, vol.~20, no.~7, pp. 4513 -- 4526, Jul. 2021.

\bibitem{Ref_niu2022double}
H.~Niu \emph{et~al.}, ``Double intelligent reflecting surface-assisted
  multi-user {MIMO} mmwave systems with hybrid precoding,'' \emph{{IEEE} Trans.
  Veh. Technol.}, vol.~71, no.~2, pp. 1575 -- 1587, Feb. 2022.

\bibitem{Ref_jiang2022intelligent}
W.~Jiang and H.~Schotten, ``Intelligent reflecting vehicle surface: A novel
  {IRS} paradigm for moving vehicular networks,'' in \emph{Proc. 2022 IEEE 40th
  Military Commun. Conf. (MILCOM 2022)}, Rockville, MA, USA, Nov. 2022.

\bibitem{Ref_tse2005fundamentals}
D.~Tse and P.~Viswanath, \emph{Fundamentals of Wireless Communication}.\hskip
  1em plus 0.5em minus 0.4em\relax Cambridge, United Kingdom: Cambridge
  University Press, Sep. 2005.

\bibitem{Ref_jiang2016robust}
W.~Jiang, T.~Kaiser, and A.~J.~H. Vinck, ``A robust opportunistic relaying
  strategy for co-operative wireless communications,'' \emph{{IEEE} Trans.
  Wireless Commun.}, vol.~15, no.~4, pp. 2642--2655, Apr. 2016.

\bibitem{cvx}
M.~Grant and S.~Boyd, ``{CVX}: Matlab software for disciplined convex
  programming, version 2.1,'' \url{http://cvxr.com/cvx}, Mar. 2014.

\end{thebibliography}

\end{document}